# Robotic Astronomy with the Faulkes Telescopes and Las Cumbres Observatory Global Telescope

Fraser Lewis[1,2,3], Rachel Street[3], Paul Roche[1,2,3], Vanessa Stroud[1,2,3], David M. Russell[4]

[1] Faulkes Telescope Project, School of Physics and Astronomy, 5, The Parade, Cardiff, CF24 3AA, UK

[2] Department of Physics and Astronomy, The Open University, Milton Keynes, MK7 6AA, UK

[3] Las Cumbres Observatory Global Telescope, 6740 Cortona Drive, Suite 102, Goleta, CA 93117, USA

[4] Astronomical Institute 'Anton Pannekoek', University of Amsterdam, P.O. Box 94249, 1090 GE, Amsterdam, the Netherlands

We present results from ongoing science projects conducted by members of the Faulkes Telescope (FT) team and Las Cumbres Observatory Global Telescope (LCOGT). Many of these projects incorporate observations carried out and analysed by FT users, comprising amateur astronomers and schools.

We also discuss plans for the further development of the LCOGT network.

## 1. Introduction

The Faulkes Telescope Project (www.faulkes-telescope.com) is based in Cardiff, Wales, and is an educational and research arm of LCOGT (http://lcogt.net), based in Goleta, California. Presently, the network has two operational 2-metre optical telescopes, located at Haleakala on Maui, Hawaii (FT North) and Siding Spring in New South Wales, Australia (FT South).

The FT provides telescope time free of charge to all UK schools, as well as other educational groups, and is able to provide its users with science and imaging projects [1]. Many of these projects relate to the UK school's curriculum, and some of them allow the schools to collaborate with researchers within the FT Team, and externally.

## 2. The Telescopes

Both FT North and FT South are 2-metre, fully autonomous, robotic Cassegrain-type reflectors with Richey-Chretien hyperbolic optics on an alt-azimuth mount. Both telescopes employ a Robotic Control System (RCS) [2].

They are currently operated in two modes:

i) UK (and some overseas) schools use a 'real-time interface' (RTI) mode, where each user has exclusive control of the telescope for 30 minutes at a time. During this time, the users are able to select co-ordinates, filters and exposure times. Many educational users participate in data-gathering for research programmes.

ii) The 'offline' mode of the telescope lends itself more easily to science observing, using a Graphic User Interface (GUI). In this instance, users may specify 'flexible' (observe when most suitable), 'fixed-time' (for specifically timed transient events) or 'monitor' (repeat these observations every few days/weeks) type observations. For either mode, the telescopes also have an over-ride capability for Target of Opportunity (ToO) events such as gamma-ray bursts.

All observations are pipeline-processed to take account of flat-fielding and de-biasing, with the finished images in 2x2 binned FITS format available typically 24 hours after the observing run.

The telescopes are equipped with a 4.6' x 4.6' field of view Merope camera [3] featuring SDSS u'g'r'i'z', Bessel BVRI, Pan-STARRS Z,Y broad-band filters in addition to H$\alpha$, H$\beta$ and O III narrow-band filters. The cameras use an E2V CCD42-40DD CCD with 2048 x 2048 pixels.

A second camera has recently been installed, known as the Spectral camera and possessing a wider field of view of 10.5' x 10.5'. The Spectral cameras have an improved readout time of ~ 5 seconds. The CCD on this camera is a Fairchild CCD 486 with 4096 x 4096 pixels.

It is planned that all future telescopes within the LCOGT network (i.e. 0.4m and 1.0m) will have similar filter sets.

## 3. Scientific Results

### 3.1 LMXB Monitoring. The Black Hole X-ray Binary, GX339-4

X-ray binaries (XRBs) consist of a pair of stars orbiting each other, however unlike 'normal' binary stars, one of the components is a compact object. This compact object is either a neutron star or a black hole. There are currently ~ 300 XRBs in the Milky Way, split into two broad categories based on the mass of the normal component of the system, namely Low-Mass and High-Mass (LMXBs and HMXBs).

XRBs possess accretion discs of matter (at temperatures of up to millions of Kelvin), which has been drawn from the 'normal' donor star. As well as supernovae, XRBs are also closely linked with phenomena such as Gamma Ray Bursts (GRBs) and as potential sources of gravitational waves.

Initial detections of XRBs are usually made at X-ray wavelengths, a signature of the highly energetic processes involved. They are also observed at other wavelengths, such as the optical regime that the FTs operate within. They can often increase in optical brightness by factors of 1000 or more in a matter of days (known as outbursts) and often display variability on the order of seconds (flickering or flaring).

The low-mass systems (LMXBs) form around 50% of the overall population - they contain a compact object and a K or M type donor star of up to ~ 1 solar mass.

Since early 2006, we have been observing a subset of 30 LMXBs with the two FTs [4]. The telescopes are an ideal vehicle with which to monitor these objects, since they allow us the flexibility to change both the frequency and intensity of our observing, subject to changes in the behaviour of our targets. No other survey of LMXBs has covered this many systems so comprehensively or over this time period.

The initial aims of the project are twofold:

i) We wish to identify outbursts. It is known that many of these systems show increases in brightness at optical wavelengths before being detected in X-rays. Associated with this is the fact that our monitoring is often more sensitive to the initial phase of outburst than typical X-ray monitoring instruments. Identifying these outbursts before other observers allows us to alert the wider astronomical community to initiate multi-wavelength follow-up observations [e.g. 5, 6].

ii) Secondly, most XRBs spend a larger fraction of their time as 'quiescent' systems rather than in outburst. We wish to constrain this duty cycle and also to compare behaviour in these two states.

On several occasions, we have been able to involve schools and other users in helping us to collect and analyse data. This has led to the schools' contribution being acknowledged in journal papers [e.g. 7, 8]. As well as contributing to professional science, studying XRBs allows students to gain a better understanding of the life cycle of stars, as studied in the science curriculum for schools in the UK.

For systems in outburst, we have been able to collect extensive datasets, incorporating supplementary data at other wavelengths from colleagues, including observatories such as the Very Large Telescope, the Very Large Baseline Array and the Swift and RXTE X-ray satellites.

One notable target for FT South is the transient black hole binary, GX339-4. This system went into outburst in late 2006 (~ MJD 54150), steadily declining in brightness over the following months. The outburst was noted at X-ray [9], optical/infrared [10] and radio wavelengths [11].

Our optical observations (Fig. 1) show a continuing decline in V, R and i' bands until ~ MJD 54585, at which time the source began to increase back to its previous brightest level (i' ~ 16) over ~ 80 days.

More recent observations (~ MJD 54900) show a continued brightening of the source to i' ~ 15. On several occasions, we have observed rapid flaring in the i' and V bands, with changes of 0.3 – 0.4 magnitudes in periods of 140 seconds. Our ATel [12] triggered multi-wavelength follow-ups with SALT, VLT, Swift and RXTE.

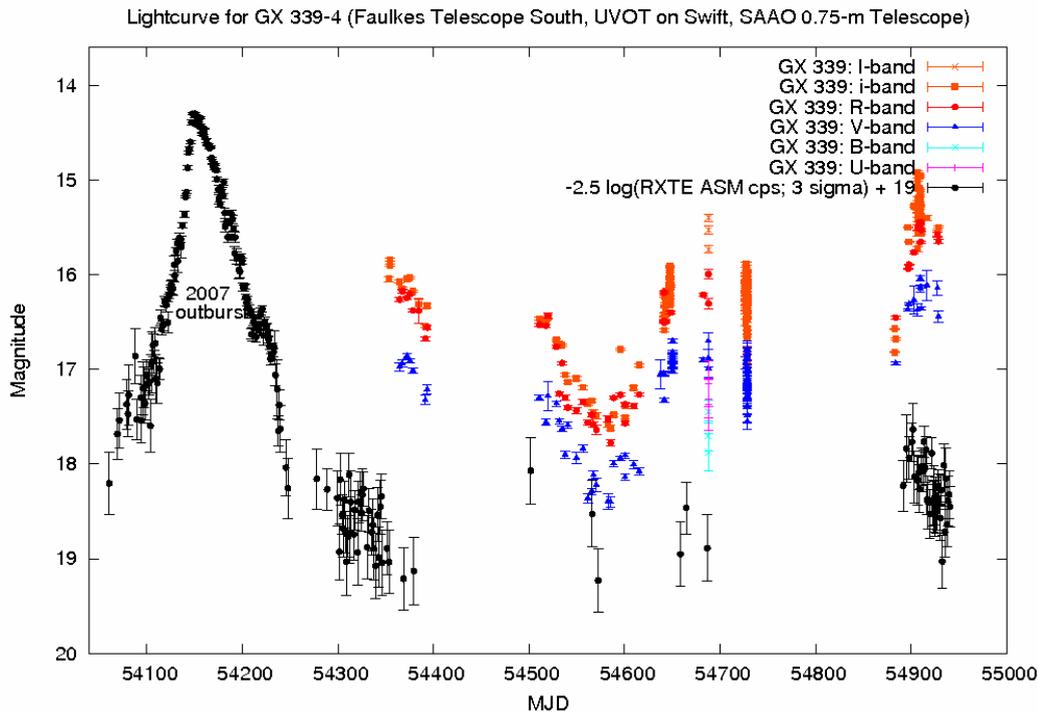

**Figure 1:** FT (V R i' ), RXTE ASM, Swift UVOT (U B) and SAAO (I) band from 2007 to present.

### 3.2 Observing Massive Stars and Open Clusters. Cyg OB2-B17

Massive stars play an important role in the structure and evolution of galaxies. They are the principal source of heavy elements and UV radiation. They are also the progenitors of energetic phenomena such as X-ray binaries and gamma-ray bursts. However, our understanding of their formation and evolution is still unclear. This is due to their short lifetimes and rapid evolutionary stages, and dust extinction making them difficult to observe during early formation phases.

Some variable massive stars include:

- OB binaries - the brightest and most massive stellar objects found in galaxies. They are the only direct way to measure the masses and radii of stars, making them the perfect test beds for studying the evolution of massive stars.

- Be Stars - rapidly rotating B stars which have circumstellar discs at some points during their lives. These are characterised by infrared excess and emission lines. How these discs are formed is still poorly understood. By studying Be stars, we can obtain information on stellar rotation and mixing.

- Luminous Blue Variables (LBVs) - massive stars close to the upper luminosity limit, causing instabilities in the stellar envelope which results in sporadic mass ejections.

We have been looking for photometric variability of massive stars in the Magellanic Clouds. This survey will be combined with multi-epoch spectroscopic data from the VLT-FLAMES study [e.g. 13].

This search has been divided into two main parts:

- Looking at lightcurves derived from the MACHO microlensing survey [14]. This survey was carried out between June 1992 and December 1999. A 50 inch telescope at Mt. Stromlo, Australia, was used with a dichroic, obtaining simultaneous blue and red observations.

- Photometric monitoring with Faulkes Telescope South. Since October 2007 we have been obtaining offline observations of the clusters NGC2004 (field covered by 5 FTS Merope camera mosaics), and N11 (8 mosaics) in the LMC and NGC346 (1 mosaic) in the SMC. A similar monitoring campaign is now monitoring 30 Doradus in the LMC, where we are obtaining observations using both offline and real-time modes. Much of the real-time data will be collected by schools.

Our aims are to look for:

- Eclipsing binaries - in order to determine the masses of the components, along with the orbital parameters.

- Variability in Be stars - gaining or losing circumstellar discs

- Variability due to sporadic mass ejections in Luminous Blue Variables

- Any transients or new phenomena within the clusters

The photometric follow-up in the Magellanic clouds is intended to reveal further systems similar to the massive binary, B17 [15].

It is an eclipsing, double-lined spectroscopic binary in the Cyg OB2 association, with two supergiants having preliminary spectral classifications of O7 and O9. The lightcurve reveals a semi-contact binary with an orbital period of $4.0217 \pm 0.0004$ days (Fig. 2) and indications of mass transfer.

B17 appears to have progenitor masses of ~ 60 solar masses, being consistent with the preliminary spectral types of both components as well as the age and stellar population of Cyg OB2. Assuming the system avoids merger, it is likely to evolve through an extreme B supergiant/LBV phase into a long period Wolf-Rayet + Wolf-Rayet binary configuration, as mass loss via stellar winds increases the orbital separation.

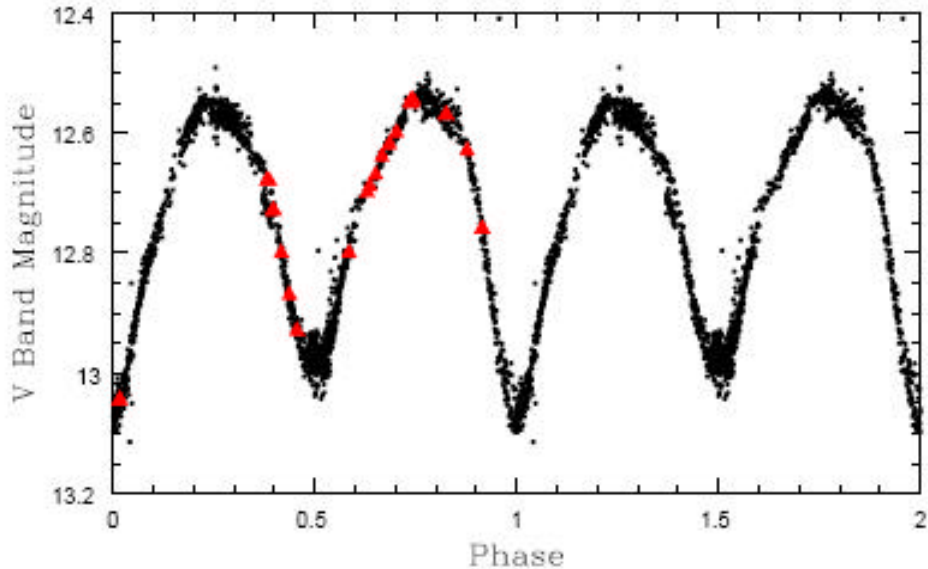

**Figure 2:** Folded lightcurve of B17 obtained by amateur astronomers. The red triangles show the phases where spectra were obtained. [15]

Open clusters are groups of tens to thousands of loosely gravitationally-bound stars, formed from the same original molecular cloud. Since the stars within a cluster essentially have the same chemical composition, age and distance, open clusters are important in the study of stellar evolution.

We have collected observations of many previously unstudied open clusters, which we encourage schools and other users to analyse. In doing this, they are able to create colour-magnitude diagrams, and search the clusters for H alpha bright objects, such as Be stars. Further, they are encouraged to image the clusters themselves to search for variable objects.

### 3.3 Microlensing Events

LCOGT supports a professional programme to observe microlensing events in the direction of the Galactic Bulge, with the aim of discovering new exoplanets. The observations we provide help to characterise the systems.

With hundreds of events every year, this program relies on a robotic system to prioritise those events which offer the most sensitivity to the presence of planets. This season, the microlensing team have established a website to provide Faulkes Telescope Project contributors with an updated list of current targets, with advice on making observations of the events using the Real-Time Interface [16]. The images taken by the schools are gathered automatically for processing and combined with data taken by astronomers around the world. This pilot program has received an encouraging response from FT users in its first year.

## 4. Future Plans

LCOGT is expanding its network to include 1-metre and 0.4 metre telescopes running in conjunction with the 2-metre FTs [17]. In addition to the imaging capability of the FTs, LCOGT are commissioning a Medium Resolution Echelle Spectrograph and a Low Resolution Spectrograph. LCOGT is designing and building a set of 1-metre telescopes, primarily for use in the LCOGT science program, but with possibly limited time available for educational use. Currently, 12-15 telescopes are planned across 6 sites, each housed in its own custom-made dome. Sites currently being planned are Siding Spring (Australia), Cerro Tololo (Chile), Mauna Loa (Hawaii), Sutherland (South Africa), Tenerife (Canary Islands) and San Pedro Martir (Mexico), with the prototype being despatched to Cerro Tololo in early 2010.

Currently in production at LCOGT headquarters in Goleta, California, the 0.4m telescopes are based around an off-the-shelf Meade telescope with a custom equatorial mount, and high specification CCD cameras. It's anticipated that these telescopes will mainly be available for the education community and other non-professional astronomers to use. Despite their relatively small size, they will be installed at dark sites in pairs or sets of fours, allowing high-quality imaging with wider fields of view (~ 30' x 20') than those of the larger aperture telescopes. They will feature the SBIG STL-6303 with a Kodak KAF 6303E CCD with 2048 x 3072 pixels.

Two of these telescopes are now located inside the enclosures of each FT. Two to four telescopes will be installed at the other 1.0m sites. When fully commissioned, they will become available to the user community. With the exception of those 0.4m telescopes housed within the FTs, the remainder of the 0.4m network will be placed within their own Aqawan enclosures, each capable of housing two telescopes [18].

### Acknowledgments
FL and VS acknowledge financial support from the Dill Faulkes Educational Trust and The Open University. DMR acknowledges support from a Netherlands Organization for Scientific Research (NWO) Veni Fellowship. ASM/RXTE data are provided by the ASM/RXTE team.

### References

[1] F. Lewis and P. Roche, "Not Just Pretty Pictures", in ".Astronomy: Networked Astronomy and the New Media", edited by R.J. Simpson, D. Ward-Thompson, 2009 (arXiv:0902.4809)
[2] Y. Tsapras et al., "RoboNet-II: Follow-up observations of microlensing events with a robotic network of telescopes", *Astronomische Nachrichten*, 330, 1, pp 4-11, 2009
[3] http://lcogt.net/en/network/camera/merope-llc-merope
[4] F. Lewis et al., "Continued monitoring of LMXBs with the Faulkes Telescopes", published in Proceedings of the 7[th] Microquasar Workshop: Microquasars and Beyond, 2008, (arXiv:0811.2336)
[5] F. Lewis et al., "Renewed optical and X-ray activity in IGR J00291+5934", *ATel*, 1726, 2009
[6] D.M. Russell and F. Lewis, "Optical and hard X-ray detections of an outburst from Aquila X-1", *ATel*, 1970, 2009

[7] P. Elebert et al., "Optical spectroscopy and photometry of SAX J1808.4-3658 in outburst", *MNRAS*, 395, pp 884 - 894, 2009

[8] M.R. Burleigh et al., "The nature of the close magnetic white dwarf and probable brown dwarf binary SDSS J121209.31+013627.7", *MNRAS*, 373, pp 1416 - 1422, 2006

[9] H.A. Krimm et al., "Swift-BAT detects a bright hard X-ray outburst from GX339-4", *ATel*, 968, 2006

[10] M. Buxton and C. Bailyn, "Latest optical and infrared observations of GX 339-4", *ATel*, 1027, 2007

[11] S. Corbel et al., "ATCA radio observations of GX339-4", *ATel*, 1007, 2007

[12] D.M. Russell et al., "Unusual optical and X-ray flaring activity in GX339-4", *ATel*, 1586, 2008

[13] C.J. Evans et al., "The VLT-FLAMES survey of massive stars: observations centered on the Magellanic Cloud clusters NGC 330, NGC 346, NGC 2004, and the N11 region", *A&A*, 456, pp 623 - 638, 2006

[14] C. Alcock et al., "Calibration of the MACHO Photometry Database", *PASP*, pp 1539 – 1558, 1999

[15] V.E. Stroud et al., "Discovery of the eclipsing Of supergiant binary, Cyg OB2-B17", *A&A*, submitted, 2009

[16] http://microlensing.lcogt.net/

[17] M. Hidas et al., "Las Cumbres Observatory Global Telescope: A homogeneous telescope network", *Astronomische Nachrichten*, 329, 3, pp 269 -270, 2008

[18] http://lcogt.net/en/blog/egomez/how-keep-telescope-dry